\documentstyle[12pt]{article}
\begin{document}

\begin{center}
\large{
A Next-to-Minimal model of Hybrid Inflation \\ in Effective
Supergravity Theories\footnote{Talk given at {\it International
Workshop on Particle Physics and the Early Universe}, Monterey,
California, 15-20 Nov 1998.}  } 
\end{center}
\begin{center}
M. Bastero-Gil and S. F. King \\
\vskip 0.5cm 
{ \it Department of Physics and Astronomy, University of
Southampton, \\
Southampton, SO17 1BJ, U. K. }
\end{center} 


\begin{abstract}
{We propose a model of inflation based on a simple variant of the
NMSSM, called $\phi$NMSSM where the additional singlet $\phi$ plays
the role of the inflaton in hybrid (or inverted hybrid) type
models. The $\phi$NMSSM solves the $\mu$
problem of the MSSM via the vacuum expectation value of the gauge
singlet N, but also solves  
the strong CP problem through an approximate Peccei-Quinn symmetry. 
The potential energy which drives
inflation originates from the F-term of the effective supergravity
theory which result from a generic string theory. In the class of
models considered  the inflaton is protected from receiving mass
during inflation by a Heisenberg symmetry, avoiding the $\eta$
problem.} 
\end{abstract}

\section*{Introduction}
Due to its intrinsic elegance, inflation \cite{inflation} has become the 
almost universally accepted dogma for 
accounting for the flatness and homogeneity of the universe.
One of the most popular versions of inflation  these days is hybrid
inflation, where there are (at least) two fields at work: 
the slowly rolling inflaton 
field $\phi$, and a second field $N$ whose role is to end inflation
by developing a non-zero vacuum expectation value (VEV) when $\phi$
passes a certain critical value $\phi_c$ during its slow roll.
During inflation $N=0$, and the potential along the $\phi$ direction is
approximately flat, with the flatness lifted by a $\phi$ mass which
must be small enough to satisfy the slow-roll conditions. 
Given a particular model, the slow roll
conditions and COBE constraints on the spectrum of density
perturbations  then determine the
relationship between the height of the potential $V(0)^{1/4}$ and the
inflaton mass during inflation. 
However, the origin of the vacuum energy $V(0)$ which drives inflation 
can only be properly understood within a framework which allows the
possibility for the potential energy to settle to zero at the
global minimum, and hence lead to an acceptable cosmological constant, and
this implies supergravity (SUGRA). In SUGRA models a non-zero vacuum energy
can be generated through the F-terms and/or the D-terms, but here we
will assume the D-terms to be negligible or zero. 

In F-term Hybrid inflation we have to face the so-called $\eta$
problem. During inflation SUSY is broken by non-zero F-terms, and 
due to the exponential factor for the Kahler potential in front of the
potential all the scalar fields including the inflaton will pick up
masses of the order of the Hubble constant, $H \approx
V(0)^{1/2}/\tilde{M}_P$. This will lead to a violation of the slow roll
condition $|\eta| = \tilde{M}_P^2 |V''/V|^2 \ll 1$. To overcome this
problem different solutions can be found in the literature
\cite{eta}. Here we would like to pursue the possibility that the
inflaton mass remains zero at tree-level during inflation and its only
contribution is given by very small radiative corrections, safely
smaller than the Hubble constant. This can be achieved working in the
context of 
no-scale SUGRA theories, where it is known that the soft scalar masses
can be zero even in the presence of a non-zero gravitino mass. Moreover,
the requirement of small inflaton mass, combined with the COBE
constraint, imply that the
height of the potential during inflation must be lower than the usual
SUSY breaking scale, $M_{SUSY} \approx 10^{11}\, GeV$, and we will
show how the no-scale structure allows this possibility
\cite{sugra}. We shall give an explicit example where the SUSY
breaking sector, the height of the  potential and the inflaton sector
are all specified.  

\section*{The model}

The model is based on the superpotential \cite{NMSSM}:
\begin{equation}
\tilde{W}=\lambda N H_1 H_2-k\phi N^2 \,,
\label{WNMSSM}
\end{equation}
with the fields $\phi$, $N$ being gauge singlets, and 
$H_1 , H_2$ the minimal supersymmetric standard model Higgs
doublets. Since the VEV of $N$ generates the effective $\mu$ mass 
term coupling the two Higgs doublets, we require that
$\lambda <N> \sim 1 $ TeV as in the well known particle physics
next-to-minimal supersymmetric standard model (NMSSM) \cite{NMSSM0}. This
superpotential is invariant under a global 
$U(1)_{PQ}$ symmetry, broken by the VEVs of $\phi$ and $N$, and which
leads to a very light  axion with its decay constant of order of the
VEVs $<N> \sim <\phi>$. This implies $<N> \sim 10^{13}\, GeV$, in
order to satisfy the cosmological axion bounds.  
The Higgs doublets develop electroweak
VEVs, much smaller than $<N>$, and they may be ignored in the analysis.

The potential relevant for inflation then reads,
\begin{equation}
V(\phi,N)= V(0) + k^2 N^4 + (m^2_N-2 k A_k \phi + 4 k^2 \phi^2) N^2 +
m_{\phi}^2 \phi^2 \, \label{Vnphi},
\end{equation}
where the soft parameters above occur in the soft SUSY breaking
potential, and they are typically of the order of $1\, TeV$, but
$m_{\phi}^2$ owes its origin to  
radiative corrections to the potential controlled by the small
coupling $k$. We have added a constant vacuum energy $V(0)$ to the
potential, whose origin we explain latter. 

For large values of the field $\phi$ the effective $N$ mass is
positive and during inflation the field $N=0$; $\phi$ slowly rolls until it
reaches a critical value\footnote{ In fact the model has two different
critical values, $\phi_c^{\pm}$, which allows the possibility of
having either standard hybrid inflation or inverted hybrid inflation
\cite{inverted} depending on the sign of the mass squared $m_{\phi}^2$.}
$\phi_c \approx \frac{A_k}{4 k}$.
When the critical value is
reached, inflation ends and the global minimum is achieved, with non
zero VEVS $<\phi>\sim <N> \sim \phi_c$.
After inflation ends $V(0)$ is assumed to remain unchanged,
but be cancelled by a negative contribution from the remaining part
of the potential at the global minimum, $V(<\phi >,<N>)= -k^2<N>^4$. 
Due to the axion bound, we then have $k \approx 10^{-10}$, and $V(0)^{1/4}
\approx 10^8\, GeV$. To satisfy the COBE constraint the model 
requires an inflaton mass in the range of a few eV, and this is consistent
with this mass being generated by radiative corrections controlled by
the small coupling\footnote{ The smallness of $k$ seems to indicate
that it has a non-renormalisable origin; this may involve an
additional sector which obeys a discrete $Z_3 \times Z_5$ symmetry
from which the Peccei-Quinn symmetry emerges as an approximation (for
details see \cite{NMSSM}).}  $k$. 

We now wish to elevate this model to an effective no-scale SUGRA
theory, where the tree-level inflaton mass is ensured to vanish. This
kind of theory may be obtained from  4d effective string theories
\cite{string}. We place the inflaton and the $N$ field in the
untwisted sector (modular weight -1) along with the moduli fields, and
assume the following conditions to hold during inflation:

(a) The superpotential is independent of the over-all modulus $T$ and,
together with Eq. (\ref{WNMSSM}),  it includes a  dilaton
superpotential $W(S)= \Lambda^3 e^{-S/b_0}$. The dilaton will act as a
source for SUSY breaking, with the gravitino mass given by:
\begin{equation}
m^2_{3/2} \simeq e^K \frac{|W(S)|^2}{\tilde{M}_P^4} \simeq
 \frac{\Lambda^6}{\tilde{M}_P^4} \,. 
\end{equation} 
The requirement of having a gravitino mass $m_{3/2} \approx 1\, TeV$
then fixes the effective scale $\Lambda$ to be of the
order of  $10^{13} GeV$ .

(b) The Kahler potential is given by,
\begin{equation}
K= -3\ln(\rho) - \ln (S+S^\ast) + \frac{\beta}{\rho^3} -\frac{2
s_0}{S+S^\ast} + \frac{b + 4 s_0^2}{6 (S+S^\ast)^2} \,, \label{Kfinal}
\end{equation}
It depends only on the combination $\rho = T + T^\ast
- \sum_i \phi_i \phi_i^\ast$, with $\phi_i$ any untwisted field of the
theory, in particular $\phi$ and $N$. This condition can be formalised in terms
of a Heisenberg symmetry \cite{heisenberg}. The twisted fields are 
switched off during inflation, and they do not contribute.
We remark that we only demand the theory to posses a Heisenberg
symmetry {\it during inflation}. After inflation ends this symmetry
may or may not be broken by the contribution from the twisted sector.   
The last three terms of Eq. (\ref{Kfinal}) model non-perturbative terms for
both the field $\rho$ and the dilaton $S$ \cite{dilaton}, needed in
order to stabilise them at a fixed value during inflation, $\rho_0 \approx
(2 \beta)^{1/3}$ and $2 Re S_0 \approx s_0$. 

The above conditions ensure that the inflaton remains massless at
tree-level. The key point is that $\rho$ is fixed 
at its minimum, for which the condition $d V/d \rho =0$ is
fulfilled. Then, computing the mass matrix for the fields ($T$,
$\phi$) during inflation it can be shown that there is a zero
eigenvalue which corresponds to the massless inflaton.

The non-zero potential $V(0)$ is  given by the SUGRA potential:
\begin{equation}
V=|F_T|^2 + |F_S|^2 - 3 m_{3/2}^2 \tilde{M}_P^2 \,, 
\end{equation}
evaluated at the minimum $\rho_0$ and $S_0$. If we now require that
the potential vanish in the 
global minimum at the end of inflation, when the fields $\phi$ and $N$
also contribute, then we obtain that during inflation the height of
the potential is 
$ V^{1/4} \sim \epsilon^{1/4} \sqrt{m_{3/2} \tilde{M}_P}$, with
$\epsilon^{1/4} \sim 10 ^{-3}$. In our approach, the fact that the
potential is much smaller than the typical SUSY breaking scale is a
consequence of having a very small coupling $k$, otherwise needed to
control the radiative corrections to the inflaton mass. Another good
feature of such a small coupling is that the contribution of the
hybrid superpotential at the end of inflation will be highly
suppressed with respect to the dilaton contribution, and then the
minima for the moduli and the dilaton (and the gravitino mass) are
mainly the same during and after inflation, i. e., no cosmological
moduli problem is present in our scenario.  As discussed in
\cite{NMSSM} the small couplings $\lambda$, $k$ may have a natural
explanation in terms of non-renormalisable effective operators. 


\end{document}